\documentclass[12pt]{article}
\usepackage{graphicx}

\def\wayne{Department of Physics\\
University of Warwick, Coventry, CV4 7AL, UK}
\def\support{\footnote{Work supported by the European Research Council.}}

\def\Title#1{\begin{center} {\Large #1 } \end{center}}
\def\Author#1{\begin{center}{ \sc #1} \end{center}}
\def\Address#1{\begin{center}{ \it #1} \end{center}}

\newenvironment{Abstract}{\begin{quotation}  }{\end{quotation}}
\newenvironment{Presented}{\begin{quotation} \begin{center} 
             PRESENTED AT\end{center}\bigskip 
      \begin{center}\begin{large}}{\end{large}\end{center} \end{quotation}}
\def\Acknowledgements{\bigskip  \bigskip \begin{center} \begin{large}
             \bf ACKNOWLEDGEMENTS \end{large}\end{center}}

\def\beq{\begin{equation}}
\def\eeq#1{\label{#1}\end{equation}}
\def\eeqn{\end{equation}}

\def\beqa{\begin{eqnarray}}
\def\eeqa#1{\label{#1}\end{eqnarray}}
\def\eeqan{\end{eqnarray}}

\let\bar=\overbar

\def\Dslash{\not{\hbox{\kern-4pt $D$}}}
\def\dslash{\not{\hbox{\kern-2pt $\del$}}}

\def\msb{{\bar{\ssstyle M \kern -1pt S}}}

\begin{document}
\begin{titlepage}
{\rightline \today}

\vfill
\Title{Open charm spectroscopy at LHCb}
\vfill
\Author{Mark Whitehead\support}
\Address{\wayne}
\vfill
\begin{Abstract}
Recent charm spectroscopy results from Dalitz plot analyses of $B$ decays to open charm final states 
at LHCb are presented. The decay modes used are $B^{+} \to D^{-} K^{+} \pi^{+}$, 
$B^{0} \to \overline{D}{}^{0} \pi^{+} \pi^{-}$ and $B^{0} \to \overline{D}{}^{0} K^{+} \pi^{-}$.
\end{Abstract}
\vfill
\begin{Presented}
The 7th International Workshop on Charm Physics (CHARM 2015)\\
Detroit, MI, 18-22 May, 2015
\end{Presented}
\vfill
\end{titlepage}
\def\thefootnote{\fnsymbol{footnote}}
\setcounter{footnote}{0}
%

\section{Introduction}

The family of charm mesons are predicted by heavy quark effective theory~\cite{godfrey} and lattice QCD~\cite{mohler}.
The 1P states have been 
well measured by the $B$-factories and LHCb~\cite{belle1,babar1,babar2,lhcb1}. 
Evidence for higher mass $D(2600)$ and $D(2760)$ states has been seen~\cite{babar2,lhcb1}. 
Only natural spin-parity resonances ($J^{P}=$ $0^{+}$, $1^{-}$, $2^{+}$,...) contribute in 
$B \to D_{(s)}hh'$ decays where $h$ and $h'$ are kaons and pions. 
In 2014 LHCb published results from a Dalitz plot analysis of $B^{0}_{s} \to \overline{D}{}^{0} K^{-} \pi^{+}$ decays,
which included the first observation of the $D_{s1}(2860)^{-}$ and $D_{s3}(2860)^{-}$ mesons~\cite{lhcb2,lhcb3}. These states 
are thought to be members of the $D_{s}$ 1D family~\cite{song,wang}.
It is therefore interesting to explore $D$ meson spectroscopy to find and identify new states to compare their properties 
with the theory predictions.
Three analyses are presented, using $B^{+} \to D^{-} K^{+} \pi^{+}$~\cite{lhcb4}, 
$B^{0} \to \overline{D}{}^{0} \pi^{+} \pi^{-}$~\cite{lhcb5} and $B^{0} \to \overline{D}{}^{0} K^{+} \pi^{-}$ decays~\cite{lhcb6}.

\section{Dalitz plot analysis of $B^{+} \to D^{-} K^{+} \pi^{+}$ decays}

\begin{figure}[!htb]
\centering
\includegraphics[scale=0.35]{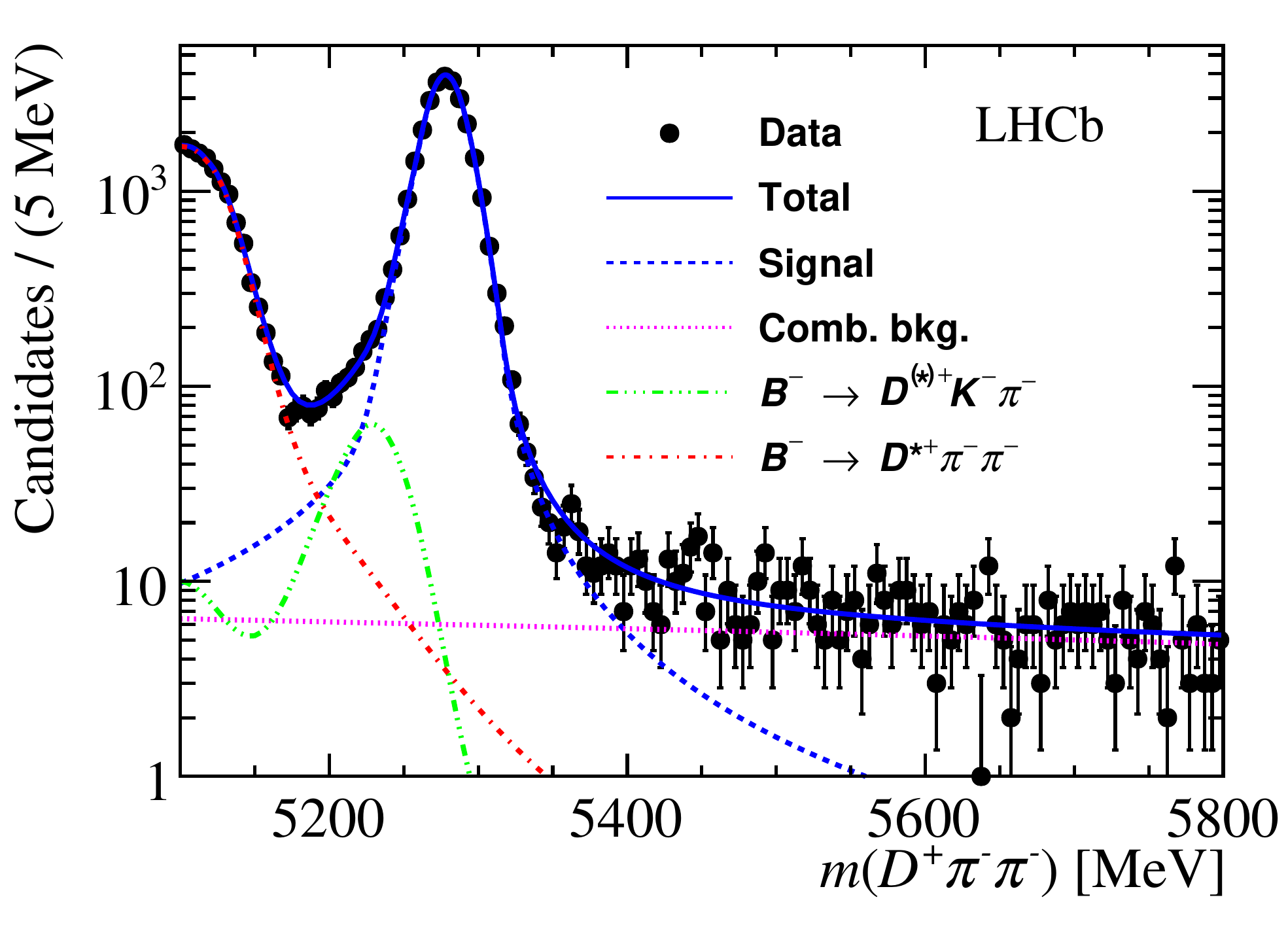} 
\includegraphics[scale=0.35]{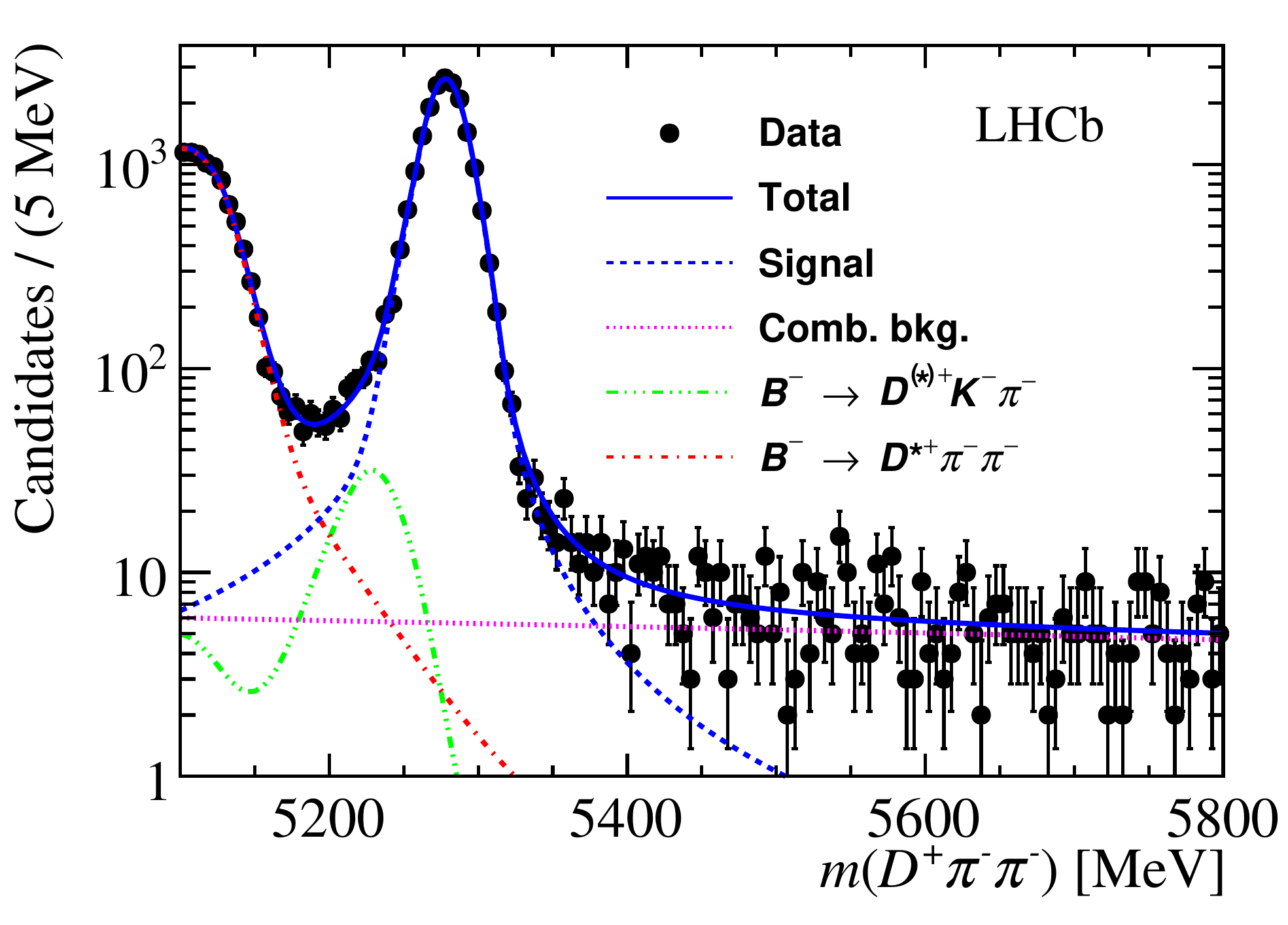} 
\caption{\small Fits to the $B$ candidate invariant mass distribution for (left) $B^{+} \to D^{-} \pi^{+} \pi^{+}$ 
and (right) $B^{+} \to D^{-} K^{+} \pi^{+}$ candidates. Components are described in the legend.}
\label{fig:1:massfit}
\end{figure}

The first observation of the decay $B^{+} \to D^{-} K^{+} \pi^{+}$, with $D^{-} \to K^{+} \pi^{-}\pi^{-}$, is made using 
the topologically similar decay $B^{+} \to D^{-} \pi^{+} \pi^{+}$ as a normalisation channel~\cite{lhcb4}. Event selection is 
based on a neural network used to reduce combinatorial background. Candidates 
in the signal and normalisation channels are shown in Fig.~\ref{fig:1:massfit},
with fits used to extract the signal and normalisation 
channel yields overlaid. Accounting for the selection efficiencies gives the branching fraction ratio
\begin{equation}
\nonumber \frac{\mathcal{B}(B^{+} \to D^{-} K^{+} \pi^{+})}{\mathcal{B}(B^{+} \to D^{-} \pi^{+} \pi^{+})} = 0.0720 \pm 0.0019 \pm 0.0021\,,
\end{equation}
where the uncertainties are statistical and systematic, respectively. Using the known value of 
$\mathcal{B}(B^{+} \to D^{-} \pi^{+} \pi^{+}) = (1.01 \pm 0.05) \times 10^{-3}$~\cite{pdg} gives
\begin{equation}
\nonumber \mathcal{B}(B^{+} \to D^{-} K^{+} \pi^{+}) = (7.31 \pm 0.19 \pm 0.22 \pm 0.39) \times 10^{-5}\,,
\end{equation}
where the third uncertainty is from $\mathcal{B}(B^{+} \to D^{-} \pi^{+} \pi^{+})$.

The Dalitz plot analysis is performed on candidates in the $B$ mass window $5239.4$--$5317.1$~MeV (natural units are used throughout), with about 
$2000$ signal candidates and a purity of approximately $93\,\%$. In $B^{+} \to D^{-} K^{+} \pi^{+}$ decays, resonances are only expected to
appear in $m(D\pi)$, allowing angular moments from the Legendre polynomials to be used to guide the amplitude model. 
The moments study showed no evidence of structures above spin 2. The 
components included in the amplitude model are shown in Table~\ref{tab:1:ampmodel}.

\begin{table}[!htb]
\centering
\begin{tabular}{lcccc}
\hline
Resonance & Spin & DP axis & Model & Parameters \\
\hline \\ [-2.5ex]
$D^{*}_{0}(2400)^{0}$ &0& $m^{2}(D\pi)$ & RBW & $m = 2318 \pm 29 {\rm MeV}$, $\Gamma = 267 \pm 40 {\rm MeV}$ \\
$D^{*}_{2}(2460)^{0}$ &2& $m^{2}(D\pi)$ & RBW & Determined from data \\
$D^{*}_{J}(2760)^{0}$ &1& $m^{2}(D\pi)$ & RBW & Determined from data \\
\hline
Nonresonant &0& $m^{2}(D\pi)$ & EFF & Determined from data \\
Nonresonant &1& $m^{2}(D\pi)$ & EFF & Determined from data \\
\hline \\ [-2.5ex]
$D^{*}_{v}(2007)^{0}$ &1& $m^{2}(D\pi)$ & RBW & $m = 2006.98 \pm 0.15 {\rm MeV}$, $\Gamma = 2.1 {\rm MeV}$ \\
$B^{*0}_{v}$ &1& $m^{2}(DK)$ & RBW & $m = 5325.2 \pm 0.4 {\rm MeV}$, $\Gamma = 0.0 {\rm MeV}$  \\
\hline
\end{tabular}
\caption{\small Components of the $B^{+} \to D^{-} K^{+} \pi^{+}$ amplitude fit model. RBW and EFF are the relativistic Breit-Wigner 
function and exponential form factor, respectively. Terms with subscript $v$ are virtual components, where the resonant pole mass 
is outside of the Dalitz plot boundary.}
\label{tab:1:ampmodel} 
\end{table}

\begin{figure}[!htb]
\centering
\includegraphics[scale=0.35]{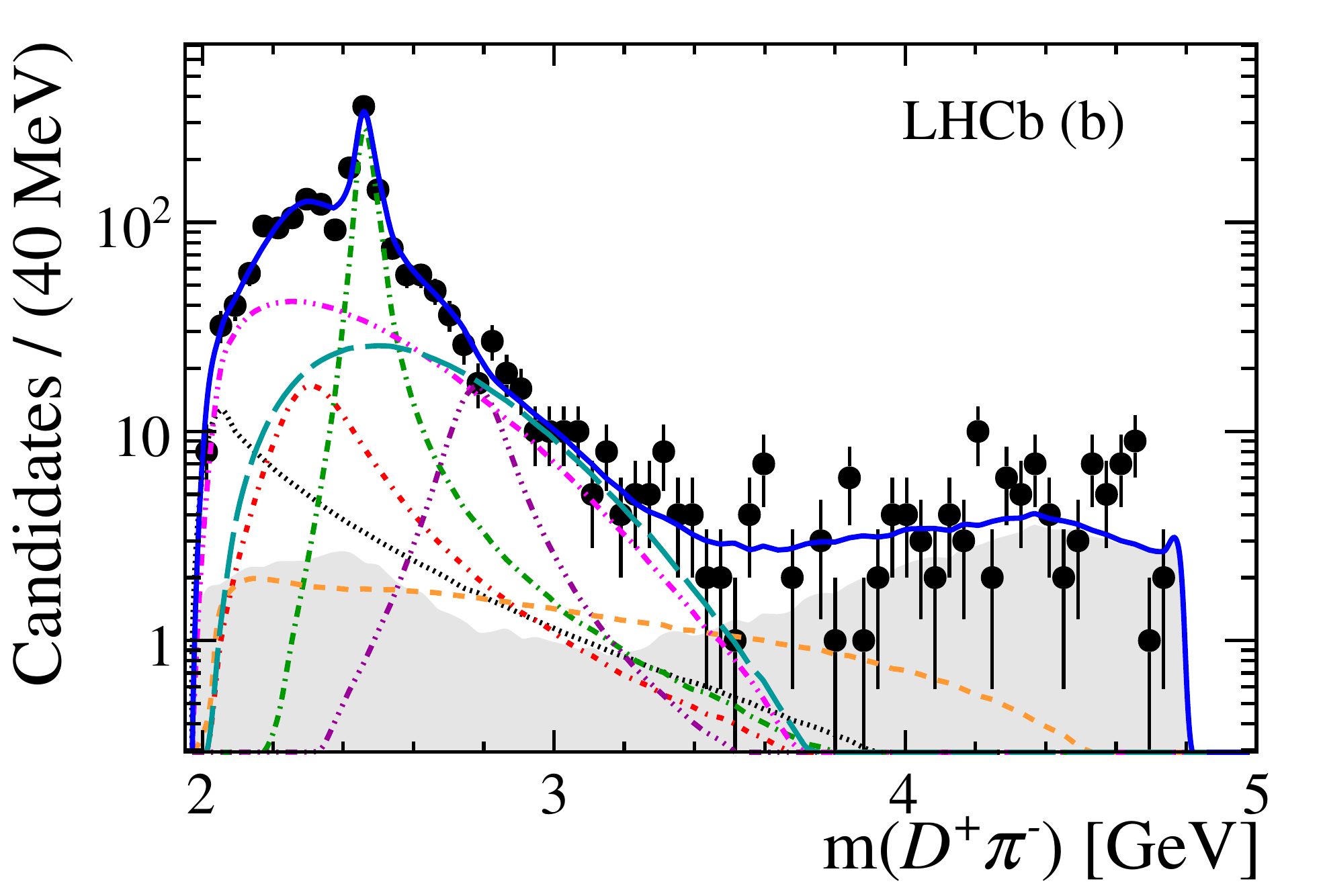} 
\includegraphics[scale=0.35,trim=0 -100 0 0,clip=true]{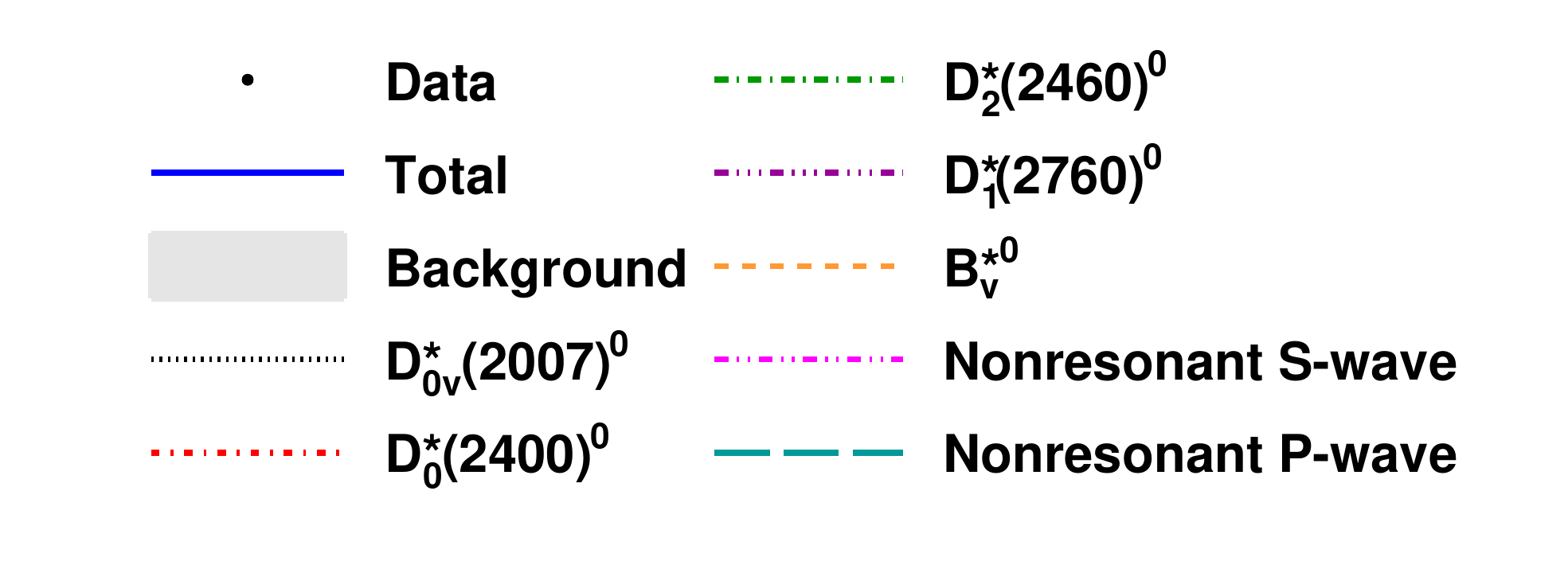} 
\caption{\small Projection of the amplitude fit in $m(D\pi)$ for $B^{+} \to D^{-} K^{+} \pi^{+}$ candidates. 
Components are as described in the legend.}
\label{fig:1:ampfit}
\end{figure}

The amplitude fit is performed with the Laura++ package~\cite{laura} using the isobar formalism~\cite{fleming,morgan,herndon}, with histograms to describe 
backgrounds and signal efficiency. For the full fit results see Ref.~\cite{lhcb4}.
Figure~\ref{fig:1:ampfit} shows the fit projection in $m(D\pi)$. Significant contributions are 
seen from the $D^{*}_{0}(2400)^{0}$, $D^{*}_{2}(2460)^{0}$ and $D^{*}_{J}(2760)^{0}$ states, 
where the spin of the latter is determined to be 1 for the first time. 
Other spin hypotheses are rejected with high significance ($>6\sigma$). 
The mass and width for the $D^{*}_{1}(2760)^{0}$ and $D^{*}_{2}(2460)^{0}$ resonances are found to be
\begin{eqnarray*}
m(D^{*}_{2}(2460)^{0})      & = & (2464.0 \pm 1.4 \pm 0.5 \pm 0.2) \,{\rm MeV} \, ,\\
\Gamma(D^{*}_{2}(2460)^{0}) & = & \phantom{24}(43.8   \pm 2.9  \pm 1.7 \pm 0.6) \,{\rm MeV} \, ,\\
m(D^{*}_{1}(2760)^{0})      & = & \phantom{.0}(2781 \pm \phantom{.}18 \pm \phantom{.}11 \pm \phantom{.0}6) \,{\rm MeV} \, ,\\
\Gamma(D^{*}_{1}(2760)^{0}) & = & \phantom{.02}(177  \pm \phantom{.}32 \pm \phantom{.}20 \pm \phantom{.0}7) \,{\rm MeV} \, ,
\end{eqnarray*}
where the uncertainties are statistical, experimental systematic and model dependent systematic, respectively.

\section{Dalitz plot analysis of $B^{0} \to \overline{D}{}^{0} \pi^{+} \pi^{-}$ decays}

\begin{figure}[!htb]
\centering
\includegraphics[scale=0.35]{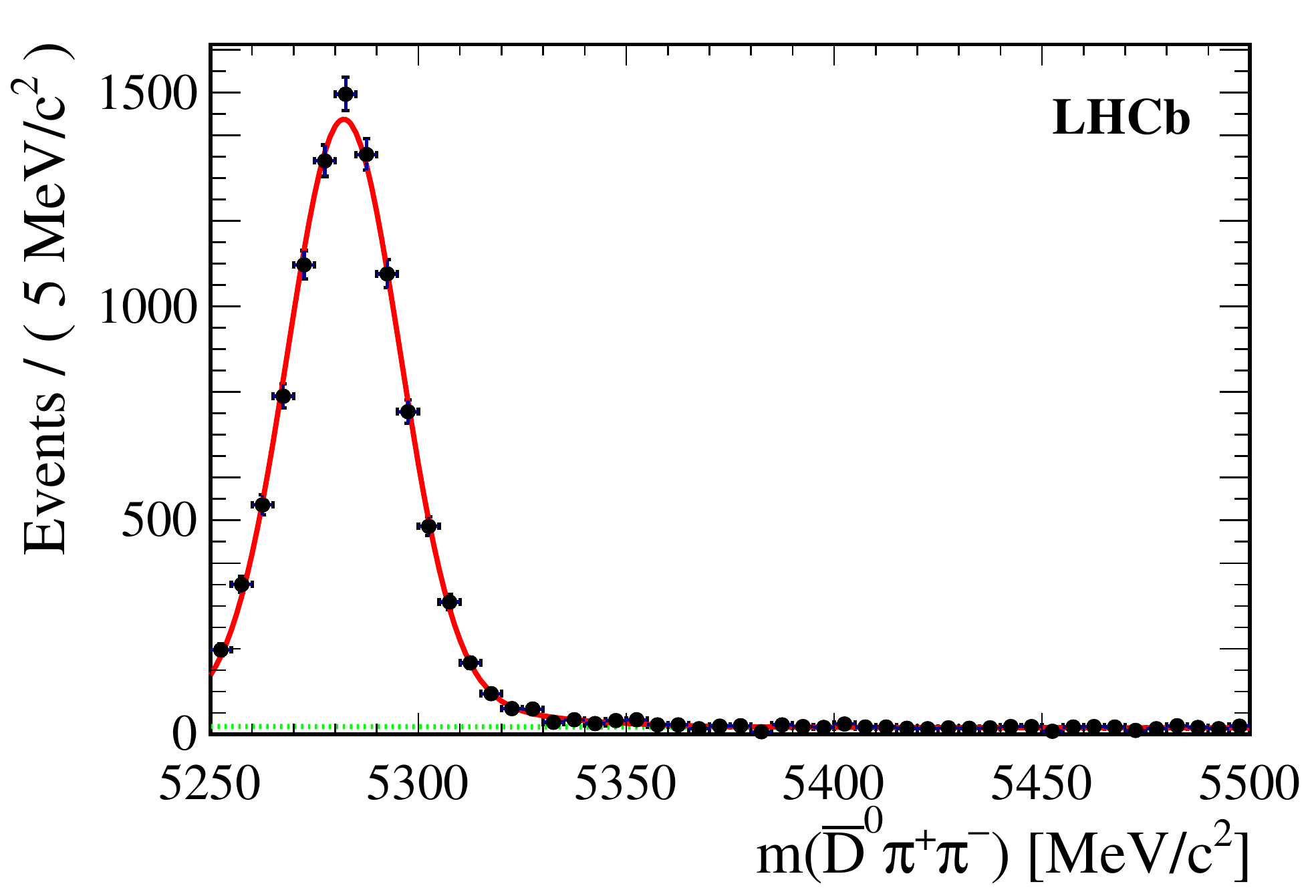} 
\caption{\small Fit to the $B$ candidate invariant mass distribution for $B^{0} \to \overline{D}{}^{0} \pi^{+} \pi^{-}$ decays.
The data are black points and the fit and backgrounds are shown in red and green, respectively.}
\label{fig:2:massfit}
\end{figure}

The amplitude analysis of the $B^{0} \to \overline{D}{}^{0} \pi^{+} \pi^{-}$ final state is performed using the $\overline{D}{}^{0}\to K^{+}\pi^{-}$ decay~\cite{lhcb5}. 
With larger data samples this channel can be used to measure $\cos(2\beta)$ and $\sin(2\beta)$~\cite{tom,beta}, where $\beta$ is an angle of the 
unitarity triangle. Resonant structures are expected in both $m(D\pi)$ and $m(\pi\pi)$. 
Combinatorial background is removed using a Fisher discriminant multivariate selection. 
Figure~\ref{fig:2:massfit} shows the $B$ candidate invariant mass distribution of selected candidates. The 
signal window used in the amplitude analysis is $5250$--$5310$~MeV. It contains about $10000$ signal candidates 
with a signal purity of around $98\,\%$.

\begin{figure}[!htb]
\centering
\includegraphics[scale=0.35]{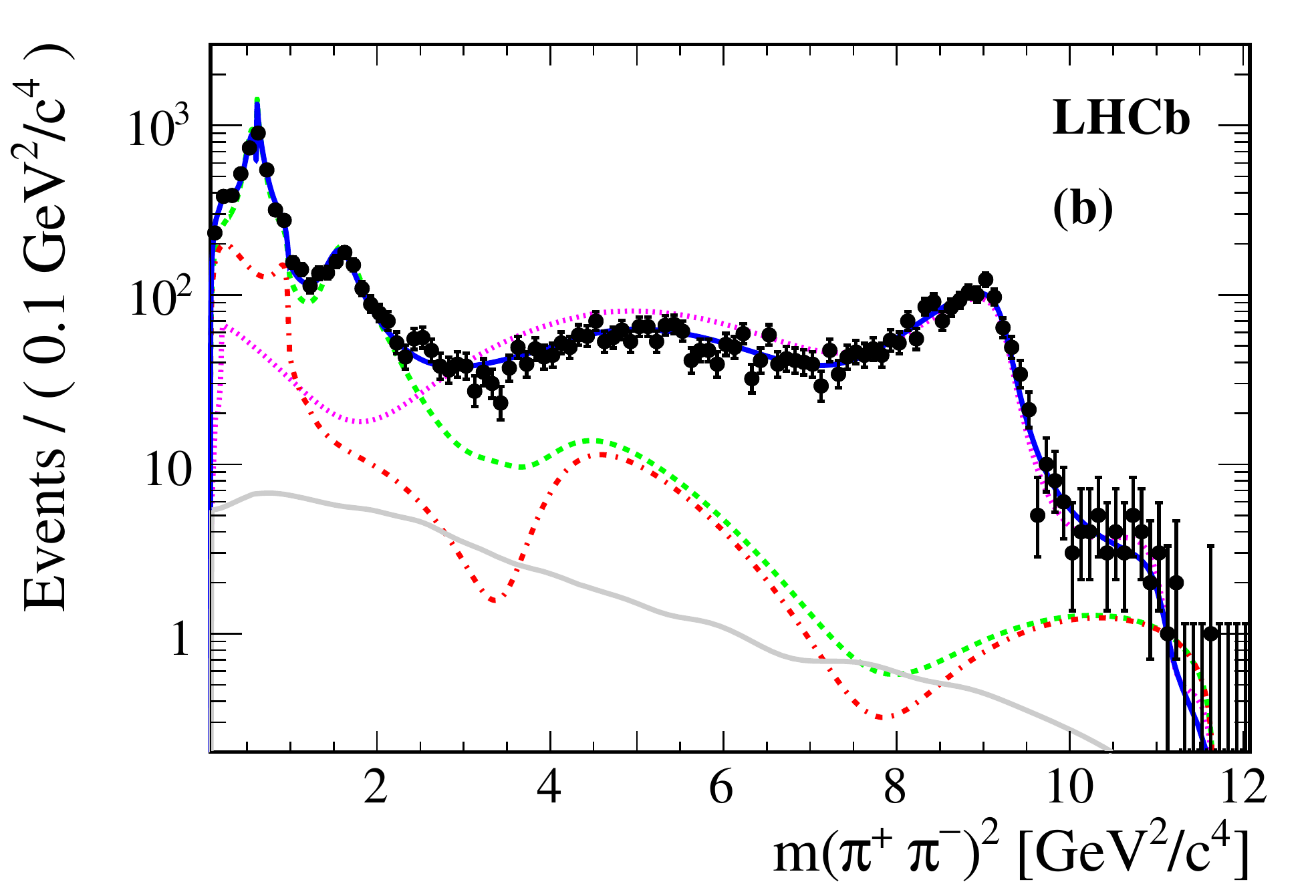} 
\includegraphics[scale=0.35]{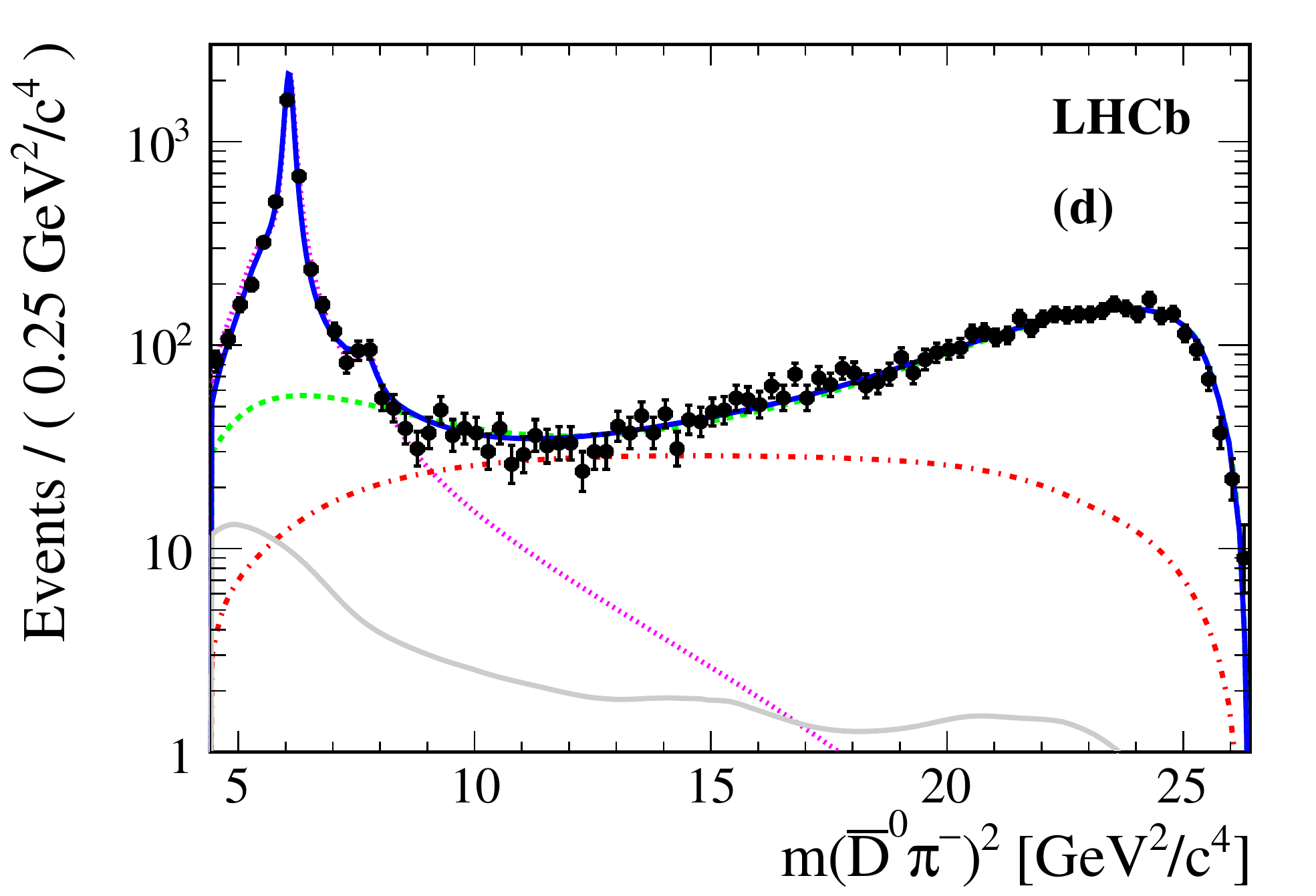} 
\caption{\small Projection of the isobar model fit in (left) $m(\pi\pi)$ and (right) $m(D\pi)$ for 
$B^{0} \to \overline{D}{}^{0} \pi^{+} \pi^{-}$ candidates. The components are (black) data, (blue) isobar fit, (green) 
$\pi^{+}\pi^{-}$ P- and D-wave, (pink) $D\pi$ contributions, (red) $\pi^{+}\pi^{-}$ S-wave and (grey) background.}
\label{fig:2:ampfit}
\end{figure}

Two amplitude fits are performed, using the isobar model and a K-matrix approach~\cite{aitchison,chung} for the $\pi\pi$ 
S-wave contribution. The resonances included in the models are shown in Table~\ref{tab:2:ampmodel}.
Projections of the isobar model fit are shown in Fig~\ref{fig:2:ampfit}, see Ref.~\cite{lhcb5} for the K-matrix results.
The charm resonances $D^{*}_{0}(2400)^{-}$, $D^{*}_{2}(2460)^{-}$ and $D^{*}_{J}(2760)^{-}$ are found 
to be significant and the $D^{*}_{J}(2760)^{-}$ state is determined, with high significance, to be spin 3 for the first time. 
It is interesting to note that in the $B^{+} \to D^{-} K^{+} \pi^{+}$ analysis the $D^{*}_{J}(2760)^{0}$ was found to be 
spin 1. This suggests that there could be two overlapping states, as was seen in the $D_s$ meson family in 
$B^{0}_{s} \to \overline{D}{}^{0} K^{-} \pi^{+}$ decays~\cite{lhcb2,lhcb3}.
The masses and widths of the charm resonances from the isobar model fit are
\begin{eqnarray*}
m(D^{*}_{0}(2400)^{-})      & = & \phantom{.6}(2349 \pm \phantom{.6}6 \pm \phantom{.0}1 \pm \phantom{.0}4) \,{\rm MeV} \, ,\\
\Gamma(D^{*}_{0}(2400)^{-}) & = & \phantom{2.6}(217   \pm \phantom{.}13  \pm \phantom{.0}5 \pm \phantom{.}12) \,{\rm MeV} \, ,\\
m(D^{*}_{2}(2460)^{-})      & = & (2468.6 \pm 0.6 \pm 0.0 \pm 0.3) \,{\rm MeV} \, ,\\
\Gamma(D^{*}_{2}(2460)^{-}) & = & \phantom{24}(47.3   \pm 1.5  \pm 0.3 \pm 0.6) \,{\rm MeV} \, ,\\
m(D^{*}_{3}(2760)^{-})      & = & \phantom{.0}(2798 \pm \phantom{.6}7 \pm \phantom{.0}1 \pm \phantom{.0}7) \,{\rm MeV} \, ,\\
\Gamma(D^{*}_{3}(2760)^{-}) & = & \phantom{.02}(105  \pm \phantom{.}18 \pm \phantom{.0}6 \pm \phantom{.}23) \,{\rm MeV} \, ,
\end{eqnarray*}
where the uncertainties are statistical, experimental systematic and model dependent systematic, respectively.
Good agreement is seen between the isobar model and K-matrix fit results.

\begin{table}[!tbh]
\centering
\begin{tabular}{lcccc}
\hline
Resonance & Spin  & Model & $m_r$ (MeV) &  $\Gamma_0$ (MeV)  \\
\hline
$\overline{D}{}^{0} \pi^-$ P-wave & 1  & ~\cite{lhcb5} & \multicolumn{2}{c}{Floated} \\
$D_0^*(2400)^-$ & 0  & RBW & \multicolumn{2}{c}{Floated} \\
$D_2^*(2460)^-$ & 2  & RBW & \multicolumn{2}{c}{Floated} \\
$D_J^*(2760)^-$ & 3  & RBW & \multicolumn{2}{c}{Floated} \\
\hline
$\rho(770)$ & 1  & GS & $775.02 \pm 0.35$ & $149.59 \pm 0.67$\\
$\omega(782)$ & 1 & ~\cite{lhcb5} & $781.91 \pm 0.24$ & $8.13 \pm 0.45$\\
$\rho(1450)$ & 1 & GS & $1493 \pm 15$ & $427\pm 31$ \\
$\rho(1700)$ & 1 & GS & $1861 \pm 17$ & $316 \pm 26$ \\
$f_2(1270)$ & 2 & RBW & $1275.1 \pm 1.2\phantom{0}$ & $185.1^{+2.9}_{-2.4}$ \\
\hline
$\pi\pi$ S-wave & 0 & K-matrix & \multicolumn{2}{c}{~\cite{lhcb5}} \\
\hline
$f_0(500)$ & 0 & ~\cite{lhcb5} & \multicolumn{2}{c}{~\cite{lhcb5}} \\
$f_0(980)$ & 0 & FLT & \multicolumn{2}{c}{~\cite{lhcb5}} \\
$f_0(2020)$ & 0 & RBW & $1992 \pm 16$ & $442 \pm 60$\\
Nonresonant & 0 & ~\cite{lhcb5} & \multicolumn{2}{c}{~\cite{lhcb5}} \\
\hline
\end{tabular}
\caption{\small Components of the isobar and K-matrix amplitude fit models to $B^{0} \to \overline{D}{}^{0} \pi^{+} \pi^{-}$ decays. 
GS is the Gounaris-Sakurai function and FLT is the Flatt\'{e} shape. See Ref.~\cite{lhcb5} for more details.}
\label{tab:2:ampmodel}
\end{table}

\section{Dalitz plot analysis of $B^{0} \to \overline{D}{}^{0} K^{+} \pi^{-}$ decays}

An amplitude analysis of $B^{0} \to \overline{D}{}^{0} K^{+} \pi^{-}$ decays with $\overline{D}{}^{0}\to K^{+}\pi^{-}$ 
is presented~\cite{lhcb6}. The goal of studying $B^{0} \to {D} K^{+} \pi^{-}$ decays is to measure the unitarity triangle 
angle $\gamma$, as outlined in Refs.~\cite{tim1,tim2}.
It can also be used to access the same charm resonances as $B^{0} \to \overline{D}{}^{0} \pi^{+} \pi^{-}$ decays,
although the available statistics are smaller. Contributions also appear in the $m^{2}(K\pi)$ axis of the Dalitz plot.

\begin{figure}[!htb]
\centering
\includegraphics[scale=0.35]{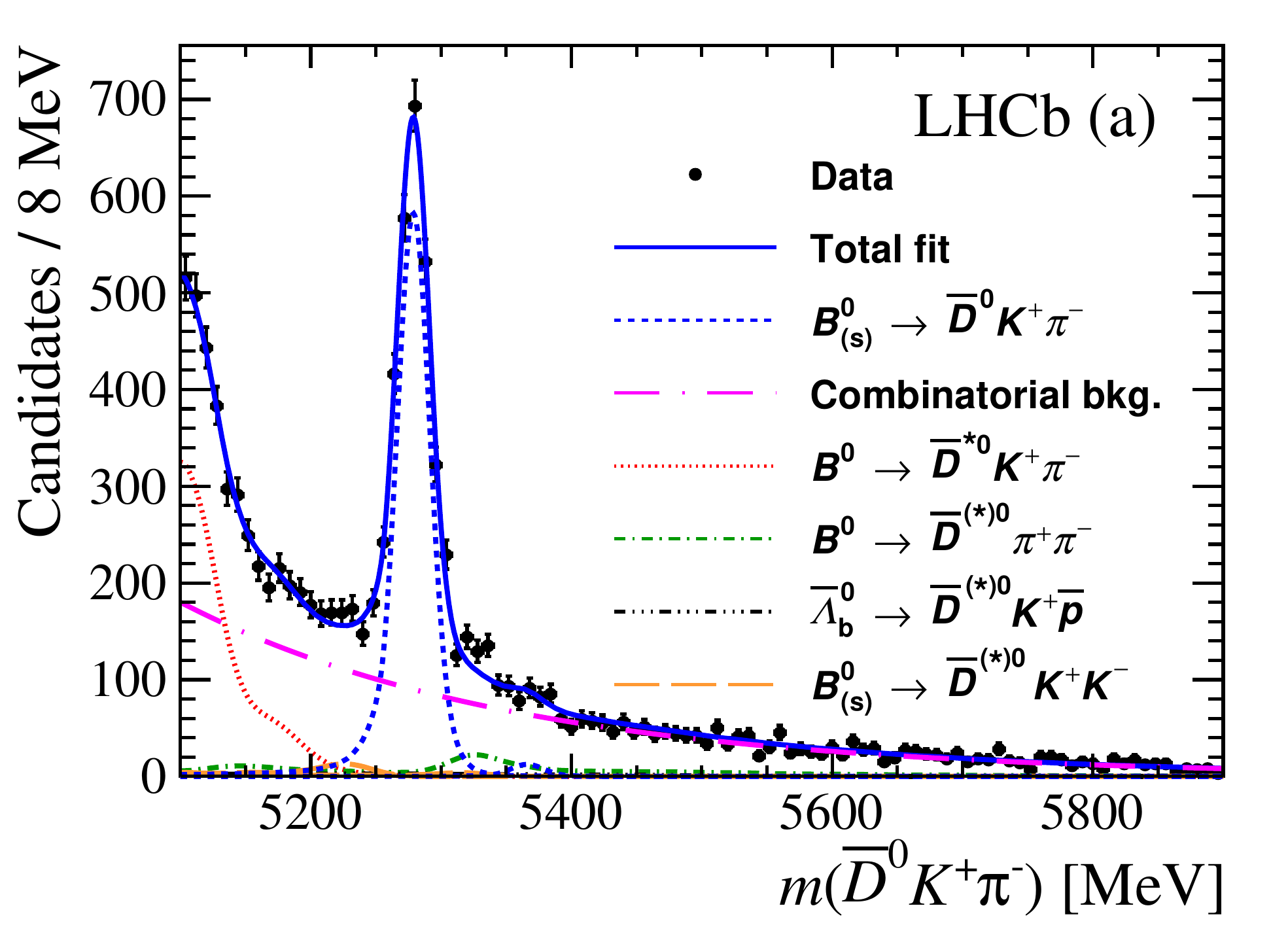} 
\includegraphics[scale=0.35]{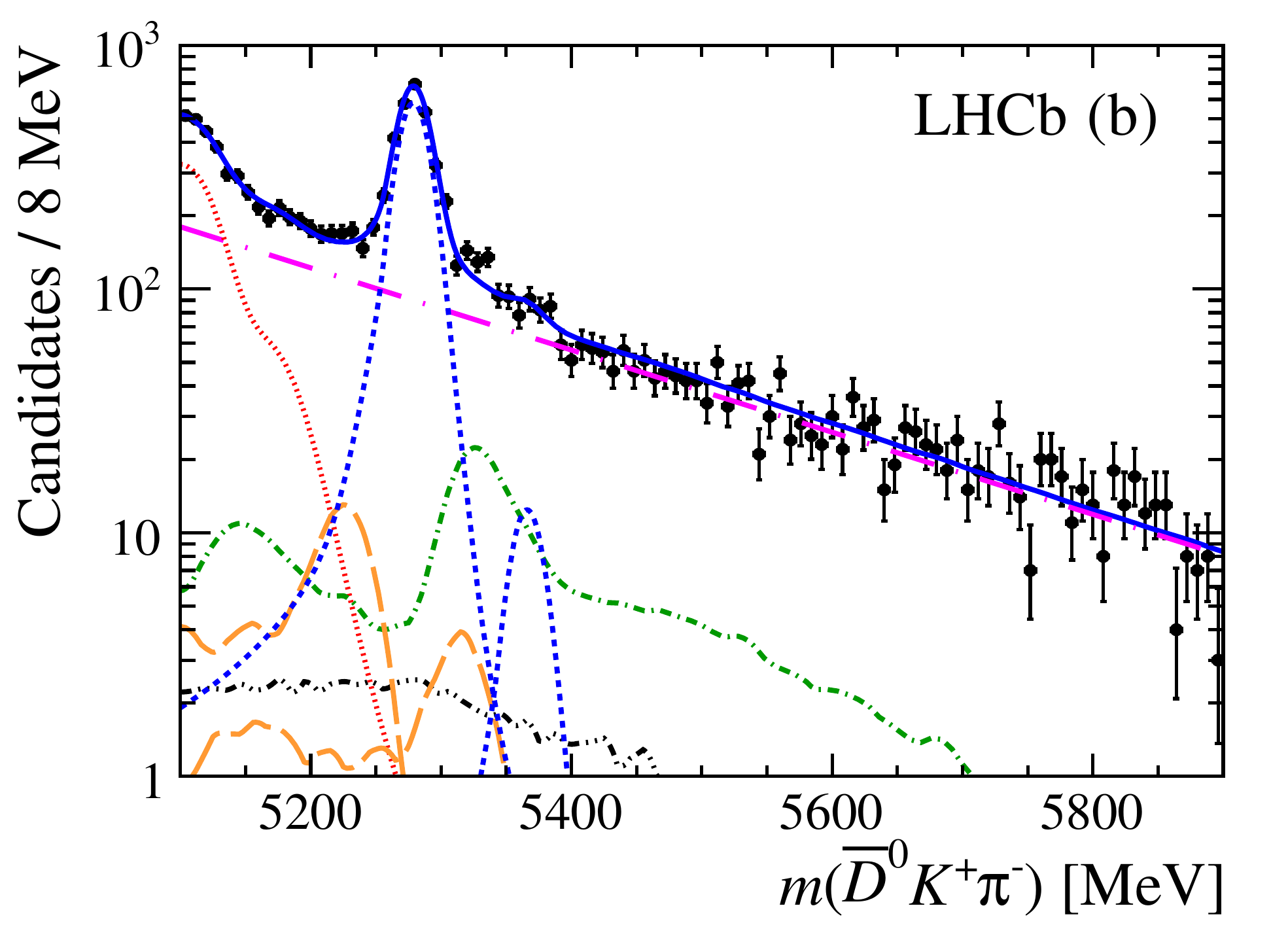} 
\caption{\small Fit to the $B$ candidate invariant mass distribution for $B^{0} \to \overline{D}{}^{0} K^{+} \pi^{-}$ candidates 
with (left) linear and (right) log $y$-axis scales. 
Components are described in the legend.}
\label{fig:3:massfit}
\end{figure}

The event selection is based on a neural network to distinguish between signal and combinatorial background. 
The $B$ candidate mass distribution of selected events is shown in Fig.~\ref{fig:3:massfit}, overlaid with the fit used to 
determine the signal and background yields. Events in the signal region, defined as $5248.6$--$5309.1$ MeV, 
are selected for the Dalitz plot fit. There are approximately $2500$ signal candidates with a purity of around 
$75\,\%$ in this window.

\begin{figure}[!htb]
\centering
\includegraphics[scale=0.35]{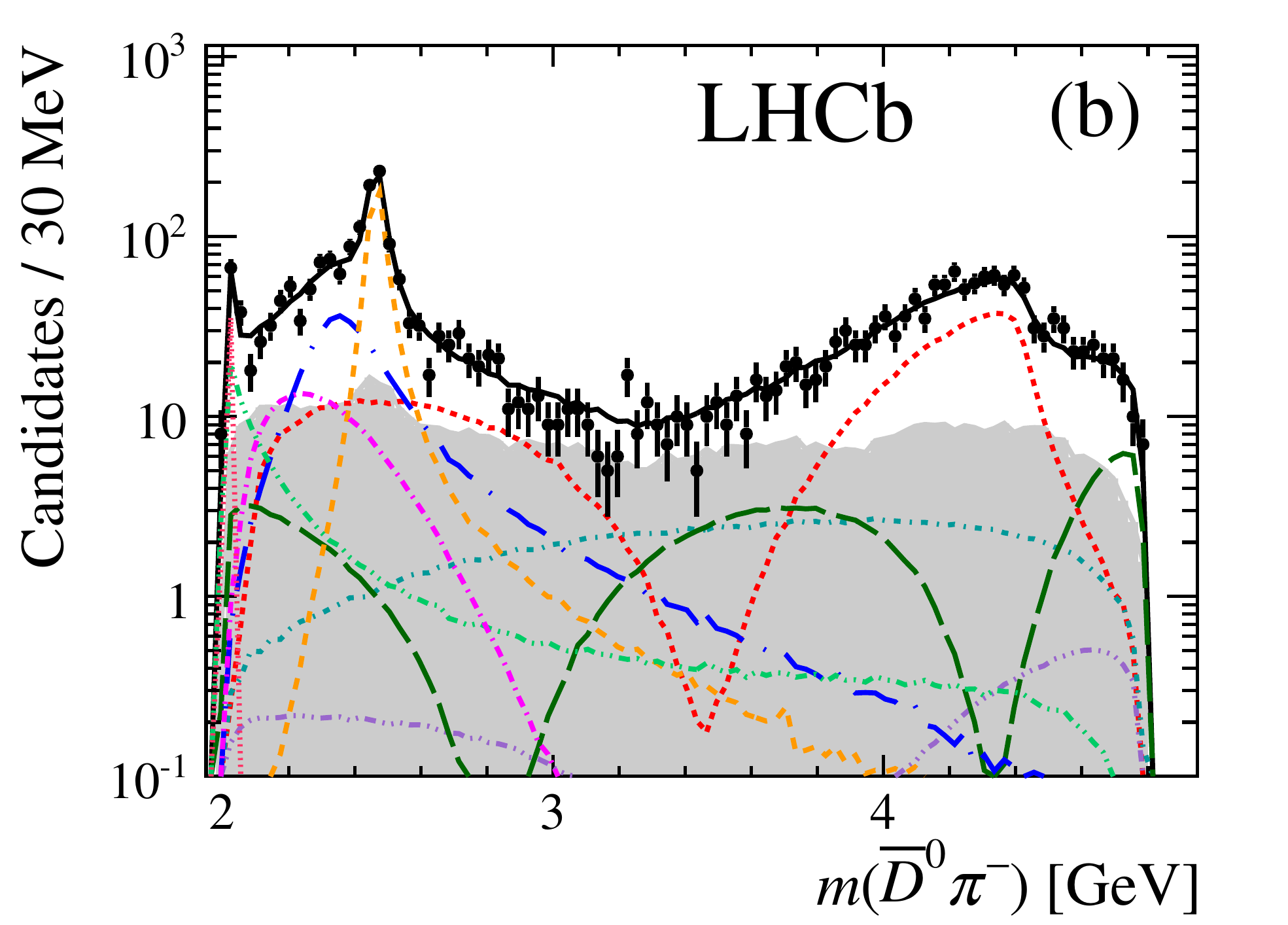} 
\includegraphics[scale=0.35]{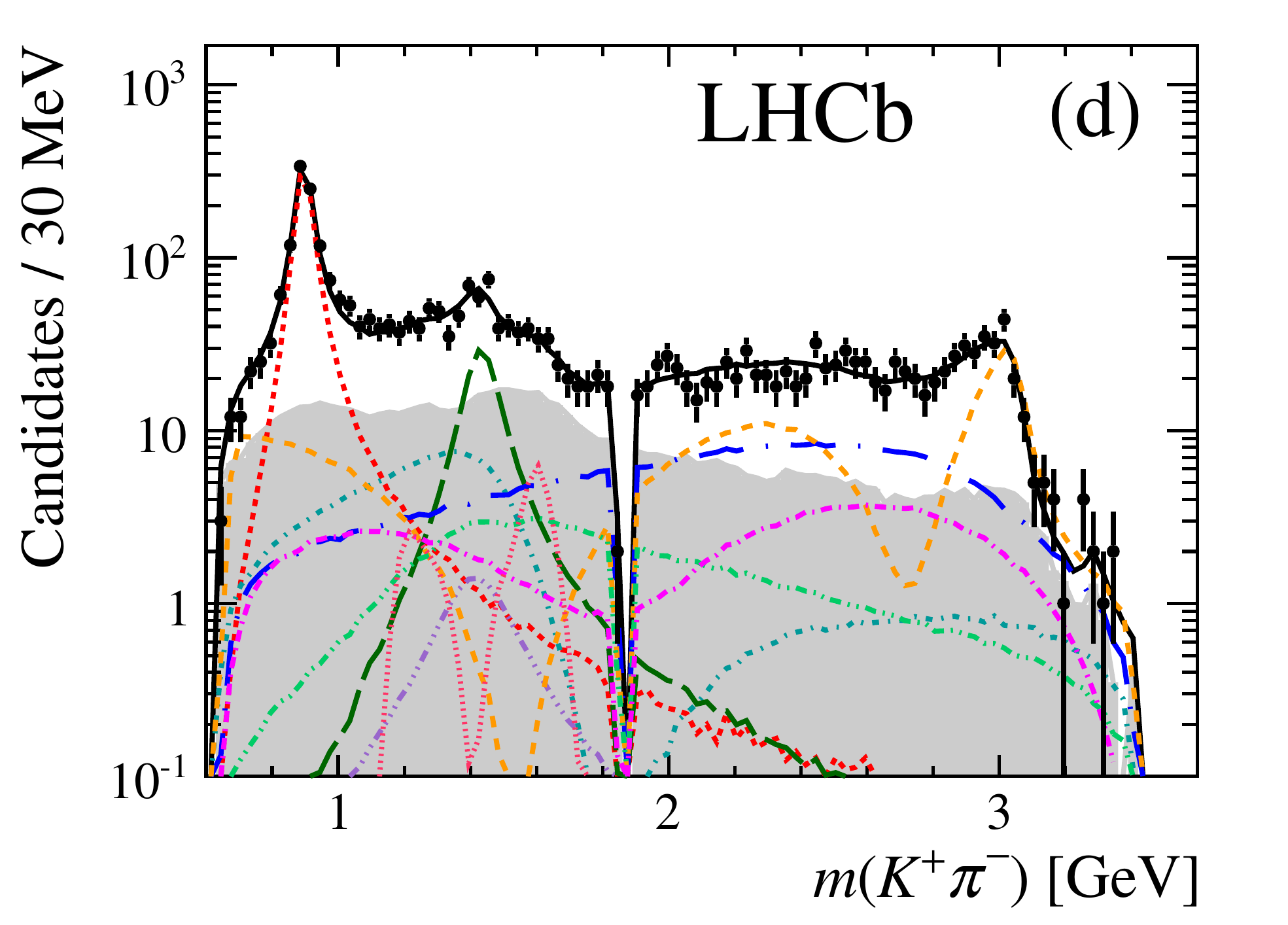} 
\includegraphics[scale=0.65]{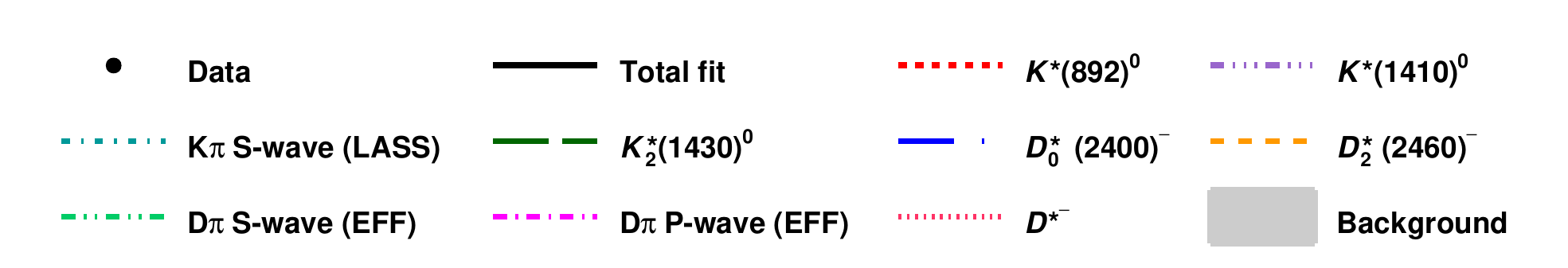} 
\caption{\small Projection of the amplitude fit in (left) $m(D\pi)$ and (right) $m(K\pi)$ for $B^{0} \to \overline{D}{}^{0} K^{+} \pi^{-}$ 
candidates. Components are described in the legend. The dip in the right plot is due to a veto at $m(D^{0})$.}
\label{fig:3:ampfit}
\end{figure}

The amplitude fit contains contributions from the terms shown in Table~\ref{tab:3:ampmodel} and
is performed using the Laura++ package~\cite{laura} with the isobar formalism. Backgrounds and efficiency corrections 
are both accounted for in the fit. For the full results of the amplitude fit see Ref.~\cite{lhcb6}. 
The projections of the amplitude fit in $m(D\pi)$ and $m(K\pi)$ are shown in Fig.~\ref{fig:3:ampfit} (left) and (right), 
respectively.

\begin{table}[!htb]
\centering
\begin{tabular}{lcccc}
\hline
Resonance & Spin & DP axis & Model & Parameters (MeV)\\
\hline \\ [-2.5ex]
$K^{*}(892)^{0}$ &1& $m^2(K\pi)$ & RBW & $m_0 = 895.81 \pm 0.19 $, $\Gamma_0 = 47.4 \pm 0.6 $ \\
$K^{*}(1410)^{0}$ &1& $m^2(K\pi)$ & RBW & $m_0 = 1414 \pm 15 $, $\Gamma_0 = 232 \pm 21 $ \\
$K^*_0(1430)^{0}$ &0& $m^2(K\pi)$ & LASS & Determined from data \\
$K^*_2(1430)^{0}$ &2& $m^2(K\pi)$ & RBW & $m_0 = 1432.4 \pm 1.3 $, $\Gamma_0 = 109 \pm 5 $ \\
$D^{*}_{0}(2400)^{-}$ &0& $m^{2}(D\pi)$ & RBW & Determined from data \\
$D^{*}_{2}(2460)^{-}$ &2& $m^{2}(D\pi)$ & RBW & Determined from data \\
\hline \\ [-2.5ex]
Nonresonant &0& $m^{2}(D\pi)$ & dabba & Fixed \\
Nonresonant &1& $m^{2}(D\pi)$ & EFF & Determined from data \\
\hline \\ [-2.5ex]
\end{tabular}
\caption{\small Components included in the $B^{0} \to \overline{D}{}^{0} K^{+} \pi^{-}$ amplitude fit model. 
More details on the dabba and LASS models can be found in Ref.~\cite{lhcb6}.}
\label{tab:3:ampmodel}
\end{table}

The charm resonance results are in agreement with the $B^{0} \to \overline{D}{}^{0} \pi^{+} \pi^{-}$ analysis. Due to the 
lower statistics available no contribution is seen at $m(D\pi) \approx 2760$ MeV. 
The masses and widths of the states $D^{*}_{0}(2400)^{-}$ and $D^{*}_{2}(2460)^{-}$ are reported to be
\begin{eqnarray*}
m(D^{*}_{0}(2400)^{-})      & = & \phantom{.0} (2360   \pm \phantom{.}15   \pm \phantom{.}12   \pm \phantom{.}28  ) \,{\rm MeV}, \\
\Gamma(D^{*}_{0}(2400)^{-}) & = & \phantom{.00} (255    \pm \phantom{.}26   \pm \phantom{.}20   \pm \phantom{.}47  ) \,{\rm MeV}, \\
m(D^{*}_{2}(2460)^{-})      & = & (2465.6  \pm 1.8  \pm 0.5  \pm 1.2 ) \,{\rm MeV}, \\
\Gamma(D^{*}_{2}(2460)^{-}) & = & \phantom{00} (46.0    \pm 3.4  \pm 1.4  \pm 2.9 ) \,{\rm MeV},
\end{eqnarray*}
where the uncertainties are statistical, experimental systematic and model dependent systematic, respectively.
These results are in agreement with those from the $B^{0} \to \overline{D}{}^{0} \pi^{+} \pi^{-}$ analysis but 
are less precise.

\section{Summary}
The latest results on charm spectroscopy from Dalitz plot analyses of $B$ meson decays at LHCb are presented. First 
observations are made of the $D^{*}_{1}(2760)^{0}$ and $D^{*}_{3}(2760)^{-}$ mesons. Larger data samples 
are needed to determine whether or not the isospin partners of these states can be seen.
%



\Acknowledgements
I thank the members of the LHCb collaboration for their help in preparing the talk and 
this document. Work supported by the European Research Council under FP7.


\end{document}